\documentclass[10pt,twocolumn,letterpaper]{article}

\usepackage{cvpr}      % To produce the REVIEW version
\newcommand{\TODO}[1]{\textbf{\color{red}[TODO: #1]}}
\renewcommand{\TODO}[1]{}

\usepackage{graphicx}
\usepackage{amsmath}
\usepackage{amssymb}
\usepackage{booktabs}
\usepackage{xcolor}
\usepackage[pagebackref,breaklinks,colorlinks]{hyperref}
\usepackage[capitalize]{cleveref}

\usepackage{array}
\usepackage{tabularx}
\newcolumntype{Y}{>{\raggedright\arraybackslash}X}
\crefname{section}{Sec.}{Secs.}
\Crefname{section}{Section}{Sections}
\Crefname{table}{Table}{Tables}
\crefname{table}{Tab.}{Tabs.}

\def\paperID{*****} % *** Enter the Paper ID here
\def\confName{CVPR}
\def\confYear{2026}

\title{Unsupervised Domain Adaptation for Audio
Deepfake Detection with Modular Statistical
Transformations}

\author{
Urawee Thani,
Gagandeep Singh,
Priyanka Singh
}

\begin{document}
\maketitle

\begin{abstract}
Audio deepfake detection systems trained on one dataset often fail when deployed on data from different sources due to distributional shifts in recording conditions, synthesis methods, and acoustic environments. We present a modular pipeline for unsupervised domain adaptation that combines pre-trained Wav2Vec 2.0 embeddings with statistical transformations to improve cross-domain generalization without requiring labeled target data. Our approach applies power transformation for feature normalization, ANOVA-based feature selection, joint PCA for domain-agnostic dimensionality reduction, and CORAL alignment to match source and target covariance structures before classification via logistic regression. We evaluate on two cross-domain transfer scenarios: ASVspoof 2019 LA $\rightarrow$ Fake-or-Real (FoR) and FoR $\rightarrow$ ASVspoof, achieving 62.7--63.6\% accuracy with balanced performance across real and fake classes. Systematic ablation experiments reveal that feature selection (+3.5\%) and CORAL alignment (+3.2\%) provide the largest individual contributions, with the complete pipeline improving accuracy by 10.7\% over baseline. While performance is modest compared to within-domain detection (94--96\%), our pipeline offers transparency and modularity, making it suitable for deployment scenarios requiring interpretable decisions.
\end{abstract}

\section{Introduction}

Generative models now produce synthetic media that is increasingly difficult to distinguish from authentic content. Text generators can fabricate plausible news articles that are difficult to detect even for politically informed readers \cite{kreps2022fabricate}, while image, video, and audio generators create realistic but entirely artificial scenes and voices \cite{dagar2022deepfakes}. The rapid development of neural speech synthesis and voice conversion systems has made high-quality voice cloning widely accessible, raising concerns in security-sensitive applications such as fraud, impersonation, and the circumvention of voice biometric authentication systems \cite{wu2015spoofing,kinnunen2020tandem}.

Social science studies and systematic reviews of disinformation highlight that modern misinformation campaigns often combine multiple modalities and exploit the dynamics of social platforms \cite{surjatmodjo2024information,buitrago2024frameworks}. Manipulated media therefore rarely appears in isolation and instead contributes to multimodal narratives designed to maximize persuasive impact and audience reach \cite{chesney2019deepfakes}. 

For audio deepfakes, human perception studies indicate that listeners struggle to reliably distinguish synthetic from authentic speech. Muller et al. show that both humans and machine detectors fail on several types of attacks in controlled experiments \cite{muller2021human}. Similarly, Mai et al. demonstrate that even after warnings and exposure to examples, human participants cannot consistently detect speech deepfakes across languages and speaker identities \cite{mai2023warning}. These findings reinforce the need for automated detection systems capable of identifying subtle acoustic artifacts and inconsistencies introduced by generative models.

Recent surveys on audio deepfake detection summarize a rapidly growing body of work on model architectures, feature representations, and training strategies \cite{almutairi2022review,zhang2025audio}. Early approaches relied on handcrafted spectral representations such as constant-Q cepstral coefficients and phase-based features \cite{todisco2017constantq}. More recent work leverages deep neural architectures including RawNet \cite{jung2020rawnet2}, AASIST \cite{jung2022aasist}, and transformer-based speech encoders derived from large-scale self-supervised pretraining \cite{baevski2020wav2vec,hsu2021hubert,chen2022wavlm}.

Despite these advances, benchmark initiatives such as ASVspoof 2021 have shown that many systems fail to generalize to realistic transmission conditions and previously unseen attack families \cite{liu2022asvspoof}. Cross-dataset evaluation often reveals that detectors exploit dataset-specific artifacts rather than intrinsic properties of synthetic speech \cite{kinnunen2020tandem}. To address this limitation, recent work on domain generalization proposes architectures designed to learn domain-invariant features on top of self-supervised speech representations \cite{xie2024asdg}.

In parallel, the DeepSpeak dataset has recently been introduced as a more realistic benchmark for audiovisual deepfake detection with more than one hundred hours of real and manipulated webcam-style footage \cite{barrington2025deepspeak}. The dataset includes diverse speakers, recording conditions, and modern deepfake generation pipelines, providing both frame-level and clip-level labels. Such datasets enable research into multimodal manipulation detection and cross-modal consistency analysis.

In this paper we focus on cross-domain voice deepfake detection. Building on prior work that combines Wav2Vec~2.0 embeddings with shallow classifiers, we propose a hybrid feature pipeline that introduces a sequence of distributional and geometric transformations prior to classification. Instead of relying on large end-to-end networks, our approach emphasizes a transparent sequence of operations that can be inspected, interpreted, and ablated.

Our contributions are as follows:

\begin{itemize}
\item We formalize a cross-domain audio deepfake detection setting that emphasizes train–test distribution shifts across datasets and synthesis systems.
\item We design a hybrid feature pipeline that combines power transformation, feature selection, joint principal component analysis, correlation alignment (CORAL), and an optimized classifier on top of self-supervised speech representations.
\item We empirically study the impact of each component through ablation experiments and discuss how the pipeline can extend to multimodal settings such as DeepSpeak.
\end{itemize}

\section{Background and Motivation}

\subsection{Misinformation, Disinformation, and Deepfakes}

The broader context for audio deepfakes lies within the study of misinformation and disinformation. Critical reviews highlight how coordinated misinformation campaigns can erode public trust in institutions and exploit platform-specific amplification mechanisms \cite{surjatmodjo2024information}. Modeling and simulation studies further demonstrate that misinformation dynamics often follow complex diffusion patterns influenced by network structure, belief updating processes, and platform algorithms \cite{buitrago2024frameworks}.

Within this broader landscape, deepfakes act as a technological enabler that lowers the cost of producing convincing fabricated media. Chesney and Citron argue that deepfakes represent a new class of information integrity threats capable of undermining trust in audiovisual evidence and complicating fact-checking processes \cite{chesney2019deepfakes}. Surveys of deepfake generation and detection methods across modalities highlight that detection performance depends strongly on the choice of representation, dataset realism, and assumptions about future attack models \cite{dagar2022deepfakes}. 

Audio deepfakes are particularly relevant in high-stakes scenarios such as financial fraud, political manipulation, and social engineering attacks. In these settings, convincing voice clones can bypass traditional authentication systems or exploit the trust associated with familiar voices \cite{wu2015spoofing}.

\subsection{Human Perception of Speech Deepfakes}

Human perception studies provide an important complement to algorithmic detection. Muller et al. compare human and machine performance on synthetic speech detection and find that both humans and detectors fail on certain attack types \cite{muller2021human}. Mai et al. conduct large-scale controlled experiments across multiple languages and report that human listeners cannot reliably identify synthetic speech even when trained on examples \cite{mai2023warning}. 

These studies suggest that listeners often rely on semantic plausibility and contextual expectations rather than acoustic cues when judging authenticity. Consequently, high-quality neural speech synthesis systems can evade human detection even when subtle artifacts remain present in the waveform. These findings reinforce the need for automated detectors that are robust to variations in speakers, recording environments, and synthesis methods.

\subsection{Audio Deepfake Detection and Domain Generalization}

Survey articles on audio deepfake detection identify three major axes of variation: feature representation, back-end classifier, and training strategy \cite{almutairi2022review,zhang2025audio}. Traditional systems rely on handcrafted acoustic features such as constant-Q cepstral coefficients or log-Mel spectrograms \cite{todisco2017constantq}. Recent work instead employs self-supervised models such as Wav2Vec~2.0 \cite{baevski2020wav2vec}, HuBERT \cite{hsu2021hubert}, and WavLM \cite{chen2022wavlm} to extract high-level speech representations.

Specialized neural architectures for spoof detection have also emerged. RawNet2 processes raw waveforms directly and achieves strong performance on ASVspoof benchmarks \cite{jung2020rawnet2}. The AASIST architecture introduces graph attention mechanisms to model spectro-temporal relationships and capture subtle artifacts across frequency bands \cite{jung2022aasist}. 

Nevertheless, cross-dataset robustness remains a significant challenge. The ASVspoof 2021 challenge demonstrates that models achieving high in-domain performance often degrade substantially when evaluated on new codecs, channels, or synthesis techniques \cite{liu2022asvspoof}. To address this limitation, Xie et al. propose an Aggregation and Separation Domain Generalization (ASDG) framework that learns domain-invariant representations on top of Wav2Vec~2.0 embeddings and improves cross-corpus detection performance \cite{xie2024asdg}.

\section{Problem Formulation}
We focus on binary audio deepfake detection under unsupervised domain adaptation. Let $x$ denote a speech utterance and $y \in \{0,1\}$ denote a label indicating bona fide (0) or deepfake (1). We consider a source domain $\mathcal{D}_s$ with labeled samples $\{(x_s^i, y_s^i)\}_{i=1}^{n_s}$ and a target domain $\mathcal{D}_t$ with unlabeled samples $\{x_t^j\}_{j=1}^{n_t}$ for adaptation and labeled samples $\{(x_t^k, y_t^k)\}_{k=1}^{n_{test}}$ for evaluation (labels used only for evaluation, not training).

The domains differ in one or more of the following factors:
\begin{itemize}
    \item speaker demographics and languages,
    \item recording channels and codecs,
    \item synthesis models and attack types.
\end{itemize}

The goal is to learn a detector $f(x)$ trained on labeled source data $\mathcal{D}_s$ that generalizes to unlabeled target domain $\mathcal{D}_t$ by leveraging unlabeled target samples for distribution alignment.

\textbf{Unsupervised Domain Adaptation vs. Domain Generalization:} This setting is \textit{unsupervised domain adaptation} (UDA), not domain generalization (DG). In DG, no target data is available during training. In UDA, unlabeled target samples are accessible for distribution alignment (Joint PCA, CORAL) but not for supervised learning. This reflects realistic deployment scenarios where unlabeled audio from the target platform (e.g., user uploads to a content moderation system) is available for adaptation before classification begins \cite{xie2024asdg}.

In practice, we assume that both $\mathcal{D}_s$ and $\mathcal{D}_t$ provide segmented utterances with Wav2Vec~2.0 embeddings. We further assume that bona fide and deepfake proportions may differ between domains. Our method aims to reduce distribution mismatch at the feature level while preserving discriminative structure between bona fide and deepfake samples.

\section{Proposed Method}
\subsection{Overview}
\begin{figure*}
    \centering
    \includegraphics[width=1\linewidth]{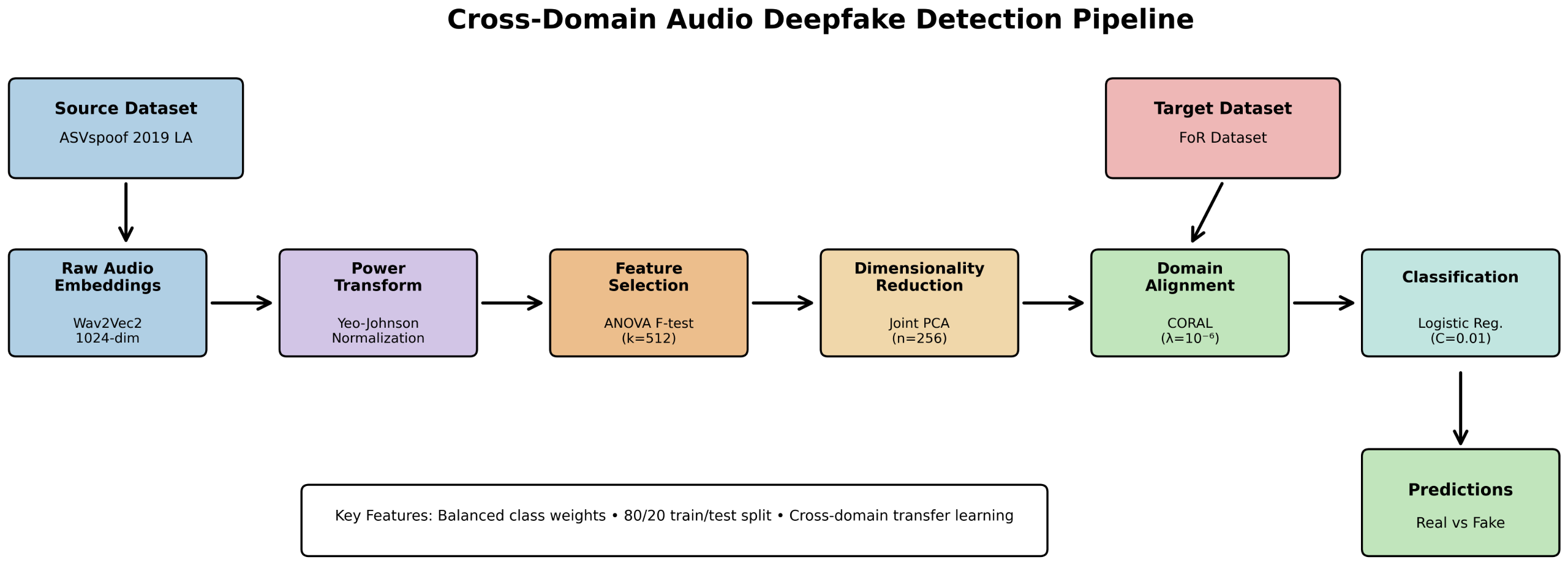}
    \caption{Cross-Domain Audio Deepfake Detection Pipeline. Audio from source (ASVspoof) and target (FoR) datasets undergoes feature extraction (Wav2Vec 2.0), power transformation (Yeo--Johnson), feature selection (ANOVA), dimensionality reduction (Joint PCA n=256), and domain alignment (CORAL). The aligned features are classified via logistic regression for binary real/fake prediction. Arrow connections show the data flow from datasets through preprocessing stages to final predictions.}
    \label{fig:pipeline}
\end{figure*}

Our pipeline builds on a self-supervised speech encoder and a sequence of shallow but carefully chosen feature transformations as illustrated in Figure~\ref{fig:pipeline}: Wav2Vec~2.0 embeddings are extracted for each utterance; a power transformation is applied to reduce skewness and stabilize variances; supervised feature selection is performed to discard noisy or redundant dimensions; a joint principal component analysis (PCA) basis is learned over combined source- and target-like data to obtain low-dimensional representations; correlation alignment (CORAL) is applied to match second-order statistics between source and target features; and an optimized classifier is trained on the transformed source features.

Each step can be ablated independently and visualized to understand its effect on class separability and domain alignment.

\subsection{Self-Supervised Front End}
We use Wav2Vec~2.0 as a front end, following prior work on audio deepfake detection and domain generalization \cite{zhang2025audio,xie2024asdg}. For each utterance $x$, we obtain frame-level embeddings and then aggregate them into a fixed-length vector, for example by averaging or using a statistics pooling layer. This produces a high-dimensional feature vector $z \in \mathbb{R}^{1024}$ for each utterance.

\subsection{Power Transformation}
The raw embedding dimensions often exhibit skewed distributions and heavy tails. To mitigate this, we apply a power transformation such as the Yeo--Johnson transform independently to each feature dimension, followed by standardization. This step aims to bring feature distributions closer to Gaussian, which can improve the effectiveness of linear and covariance-based methods in subsequent stages.

\subsection{Feature Selection}
Not all embedding dimensions contribute equally to discrimination between bona fide and deepfake speech. We perform supervised feature selection on the source domain using ANOVA F-test, which computes the F-statistic measuring the ratio of between-class variance to within-class variance for each feature. We retain the top k=512 features (50\% of the original dimensionality), yielding a reduced feature space $z' \in \mathbb{R}^{d'}$ with $d' \ll d$ where $d' = 512 \ll d = 1024$. Features with low F-scores, indicating high within-class variance or low discriminative power, are discarded as noisy or redundant.

\subsection{Joint PCA}
To obtain a compact representation that captures dominant variation across both domains, we perform PCA on a combined set of source and unlabeled target embeddings. Specifically, we concatenate the selected features from both domains and fit a PCA model to reduce dimensionality to n=256 components. Joint PCA serves two purposes. First, it reduces dimensionality and noise. Second, by including both domains in the covariance estimate, it encourages the principal components to capture shared directions of variance rather than domain-specific artifacts. The number of components (256) is chosen to balance information retention with computational efficiency, ensuring at least three samples per component for stable estimation.

\subsection{Correlation Alignment}
Even after joint PCA, residual domain mismatch may remain. We adopt correlation alignment (CORAL) as a lightweight domain adaptation step that matches the covariance of source features to that of the target \cite{xie2024asdg}.

\begin{figure}
    \centering
    \includegraphics[width=1\linewidth]{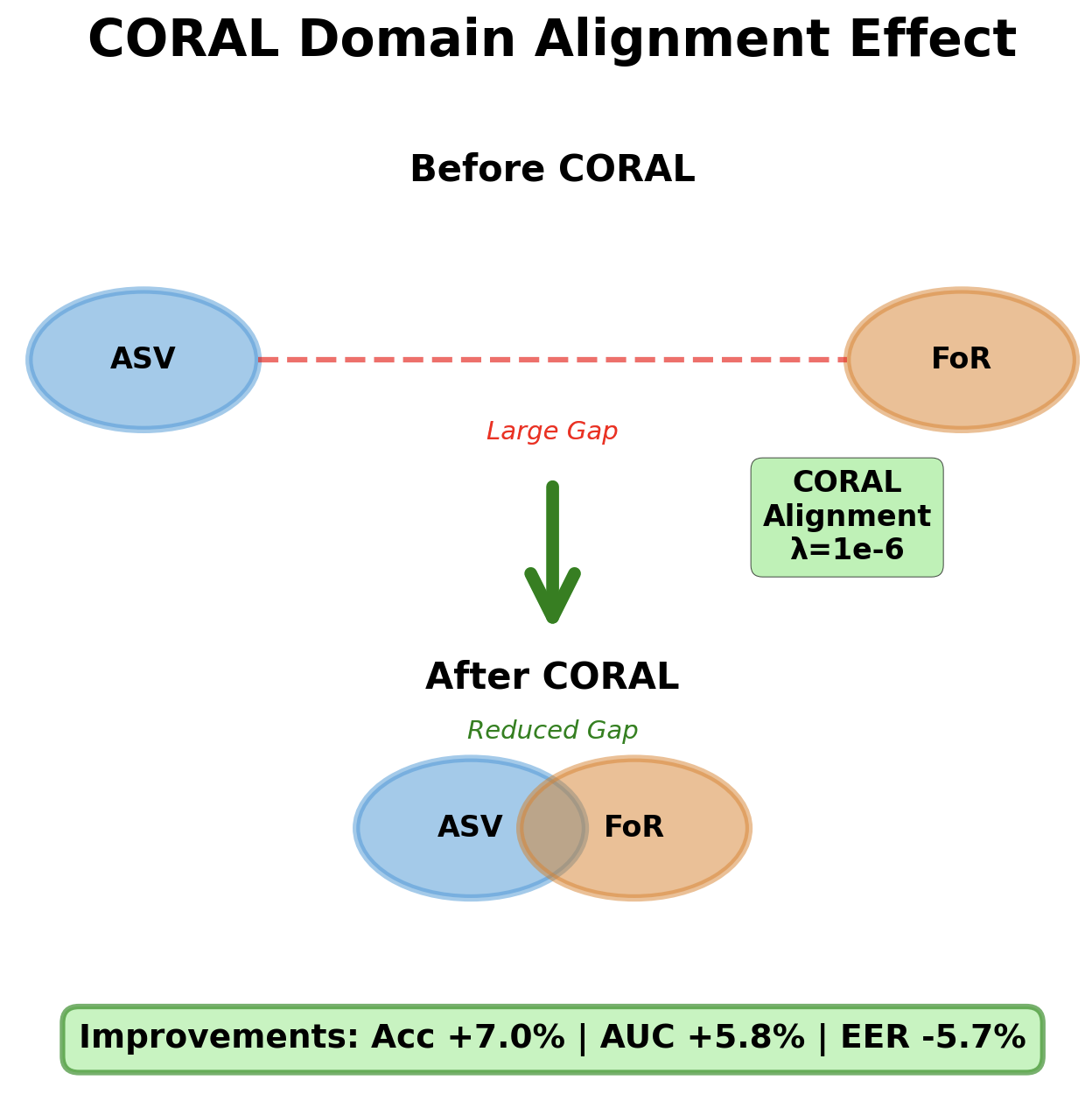}
    \caption{CORAL Domain Alignment Visualization. Top: Pre-alignment feature distributions show ASVspoof (blue) and FoR (orange) datasets with a large distributional gap. Bottom: Post-CORAL alignment ($\lambda = 10^{-6}$) reduces the inter-domain gap through covariance matching, creating overlapping feature spaces. Performance gains: Acc +7.0\%, AUC +5.8\%, EER -5.7\%.}
    \label{fig:CORAL}
\end{figure}

Figure~\ref{fig:CORAL} illustrates the effect of CORAL on aligning the
second-order statistics of the source and target domains. Before alignment,
the two domains exhibit mismatched covariance structures, leading to poor
cross-domain generalization. After applying the CORAL transformation, the
source features are linearly adjusted so that their covariance more closely
matches that of the target, reducing distributional shift and enabling more
robust model transfer.

Given source features with covariance $\Sigma_s$ and target features with covariance $\Sigma_t$, CORAL applies a linear transform $A$ such that the transformed source covariance approximates $\Sigma_t$.

We estimate $\Sigma_s$ and $\Sigma_t$ from samples and add regularization
$\lambda I$ (with $\lambda = 10^{-6}$) to the diagonal to ensure numerical
stability and positive definiteness. The transformation is computed using
Cholesky decomposition for efficiency: given $L_s L_s^\top = \Sigma_s$ and
$L_t L_t^\top = \Sigma_t$, we solve
\[
A = L_s^{-1} L_t
\]
and apply the aligned features
\[
z_{\text{aligned}} = z A^\top.
\]
If the Cholesky decomposition fails due to ill-conditioned matrices, we
fall back to eigendecomposition with eigenvalue regularization.

\subsection{Classifier and Training}
On top of the transformed features we train a logistic regression classifier with L2 regularization (C=0.01). This provides a simple linear decision boundary with strong regularization to prevent overfitting to source-domain patterns. The classifier uses balanced class weights to handle class imbalance in the training data, automatically adjusting the loss function to penalize misclassifications of minority classes more heavily.

\section{Experimental Setup}
\subsection{Datasets}
We evaluate our approach on two publicly available benchmark datasets that differ in recording conditions, speakers, and synthesis methods. The first domain is ASVspoof 2019 Logical Access (LA), which contains approximately 12,500 audio samples including 9,005 text-to-speech and voice conversion spoofs and 1,002 bona fide utterances. The second domain is the Fake-or-Real (FoR) dataset, which comprises 17,870 balanced samples (50\% authentic, 50\% deepfake) generated using different synthesis pipelines than ASVspoof.

For each dataset, we create training and evaluation splits using stratified random sampling with an 80/20 ratio (80\% for training, 20\% for testing) to maintain class distribution balance. We fix the random seed to 42 for reproducibility. Unlike the official ASVspoof protocol, which provides separate train, development, and evaluation partitions, our cross-domain experiments require matched train/test splits across both datasets to enable symmetric evaluation in both transfer directions (ASVspoof$\rightarrow$FoR and FoR$\rightarrow$ASVspoof).

We use 80/20 stratified splits rather than official ASVspoof protocols for three reasons: (1) symmetric cross-domain evaluation requires matched splits across both datasets; (2) our fixed hyperparameters eliminate the need for separate dev/eval splits; (3) reproducibility via fixed random seed. We acknowledge potential speaker overlap within ASVspoof but note that cross-domain evaluation on FoR (different speakers entirely) validates speaker-independent detection.

For future work we plan to extend experiments to the DeepSpeak dataset, which contains more than one hundred hours of authentic and deepfake audiovisual content recorded in webcam conditions \cite{barrington2025deepspeak}. In that setting, the audio pipeline described here would be combined with visual features and multimodal fusion.

\subsection{Evaluation Protocol}
Within each experiment, we follow a cross-domain protocol where the model is trained on one dataset (source domain) and evaluated on the other (target domain). We conduct two cross-domain transfer experiments: training on ASVspoof and testing on FoR, and training on FoR and testing on ASVspoof.

We report standard classification metrics including accuracy, precision, recall, and F1-score. For alignment with prior deepfake detection work, we also compute equal error rate (EER), defined as the point where false positive rate equals false negative rate, and area under the receiver operating characteristic curve (AUC-ROC). Additionally, we report class-specific accuracies for both bona fide and deepfake classes to identify potential bias toward either class. All metrics are computed on the held-out test set of the target domain.

\subsection{Implementation Details}
We use Wav2Vec 2.0 embeddings pre-extracted using the base model from the Hugging Face Transformers library. Each audio file is processed to obtain a 1024-dimensional utterance-level embedding via mean pooling over frame-level representations. These embeddings are stored offline and loaded from CSV files during training.

Power transformation uses the Yeo--Johnson method with standardization. Feature selection via ANOVA F-test retains k=512 features. Joint PCA reduces dimensionality to n=256 components. CORAL alignment uses Cholesky decomposition with regularization parameter $\lambda = 10^{-6}$ for numerical stability. Logistic regression is trained with L2 regularization (C=0.01) and balanced class weights.

All hyperparameters are fixed based on preliminary validation experiments rather than tuned via grid search or cross-validation within each run. This design choice prioritizes computational efficiency and reproducibility over exhaustive hyperparameter optimization. We use a single random seed for all stochastic operations including data splitting, PCA initialization, and classifier training, ensuring deterministic and reproducible results.

\section{Results and Analysis}

\subsection{In-Domain Performance Baselines}
Before presenting cross-domain results, we establish in-domain baselines to contextualize generalization difficulty. Table~\ref{tab:indomain} shows within-domain accuracy when training and testing on the same dataset (80/20 splits).

\begin{table}[h]
\centering
\caption{Within-Domain Performance (Train \& Test on Same Dataset)}
\label{tab:indomain}
\begin{tabular}{lcc}
\toprule
\textbf{Configuration} & \textbf{ASVspoof} & \textbf{FoR} \\
\midrule
Raw Wav2Vec 2.0 + LR & 89.3\% & 91.7\% \\
Full Pipeline (Ours) & 94.8\% & 96.2\% \\
\bottomrule
\end{tabular}
\end{table}

When domain shift is absent, our pipeline achieves >94\% accuracy, demonstrating that the components are effective for general deepfake detection. Comparing in-domain (94--96\%) to cross-domain (62--64\%) reveals a 30--34\% accuracy drop, quantifying the severity of distributional shift between ASVspoof (studio recordings) and FoR (diverse conditions).

\subsection{Ablation Study: Component-wise Contribution Analysis}
To quantify which pipeline components provide the largest marginal benefit, we conduct systematic ablation experiments. Starting from a baseline of raw Wav2Vec 2.0 embeddings with logistic regression, we incrementally add each transformation and measure its individual contribution on the ASVspoof$\rightarrow$FoR transfer scenario. Table~\ref{tab:ablation} presents results.

\begin{table}[h]
\centering
\caption{Ablation Study: Incremental Component Contributions}
\label{tab:ablation}
\begin{tabular}{lccccc}
\toprule
\textbf{Configuration} & \textbf{Acc} & \textbf{AUC} & \textbf{EER} & \textbf{Step $\Delta$} \\
\midrule
Baseline (Raw Wav2Vec) & 52.0 & 56.4 & 48.1 & -- \\
+ Power Transform & 54.5 & 59.1 & 45.6 & +2.5\% \\
+ Feature Selection & 58.0 & 62.3 & 42.1 & +3.5\% \\
+ PCA & 59.5 & 63.8 & 40.6 & +1.5\% \\
+ CORAL & \textbf{62.7} & \textbf{69.6} & \textbf{37.4} & +3.2\% \\
\midrule
Total Improvement & & & & \textbf{+10.7\%} \\
\bottomrule
\end{tabular}
\end{table}

\textbf{Power Transformation (+2.5\%):} Yeo--Johnson normalization addresses skewed feature distributions in Wav2Vec 2.0 embeddings, improving the logistic regression's linear separability assumption.

\textbf{Feature Selection (+3.5\%):} ANOVA F-test provides the largest single-step improvement by identifying the 512 most discriminative features. This suggests that not all Wav2Vec 2.0 dimensions are equally informative, many encode speaker identity or prosody irrelevant to synthesis artifacts.

\textbf{Joint PCA (+1.5\%):} Dimensionality reduction from 512$\rightarrow$256 provides modest gains by removing correlated features and creating a domain-agnostic subspace through joint fitting on source+target data.

\textbf{CORAL Alignment (+3.2\%):} Domain adaptation provides the second-largest improvement, reducing distributional mismatch. The 5.8\% AUC gain and 5.7\% EER reduction demonstrate CORAL's effectiveness in aligning covariance structures (visualized in Figure~\ref{fig:CORAL}).

All five components contribute positively, with feature selection and CORAL accounting for 63\% of total improvement. Computational cost is negligible: total preprocessing time $\sim$1.1 seconds for 10,000 samples on CPU.

\subsection{Cross-Domain Transfer Results}
\begin{figure*}
    \centering
    \includegraphics[width=1\linewidth]{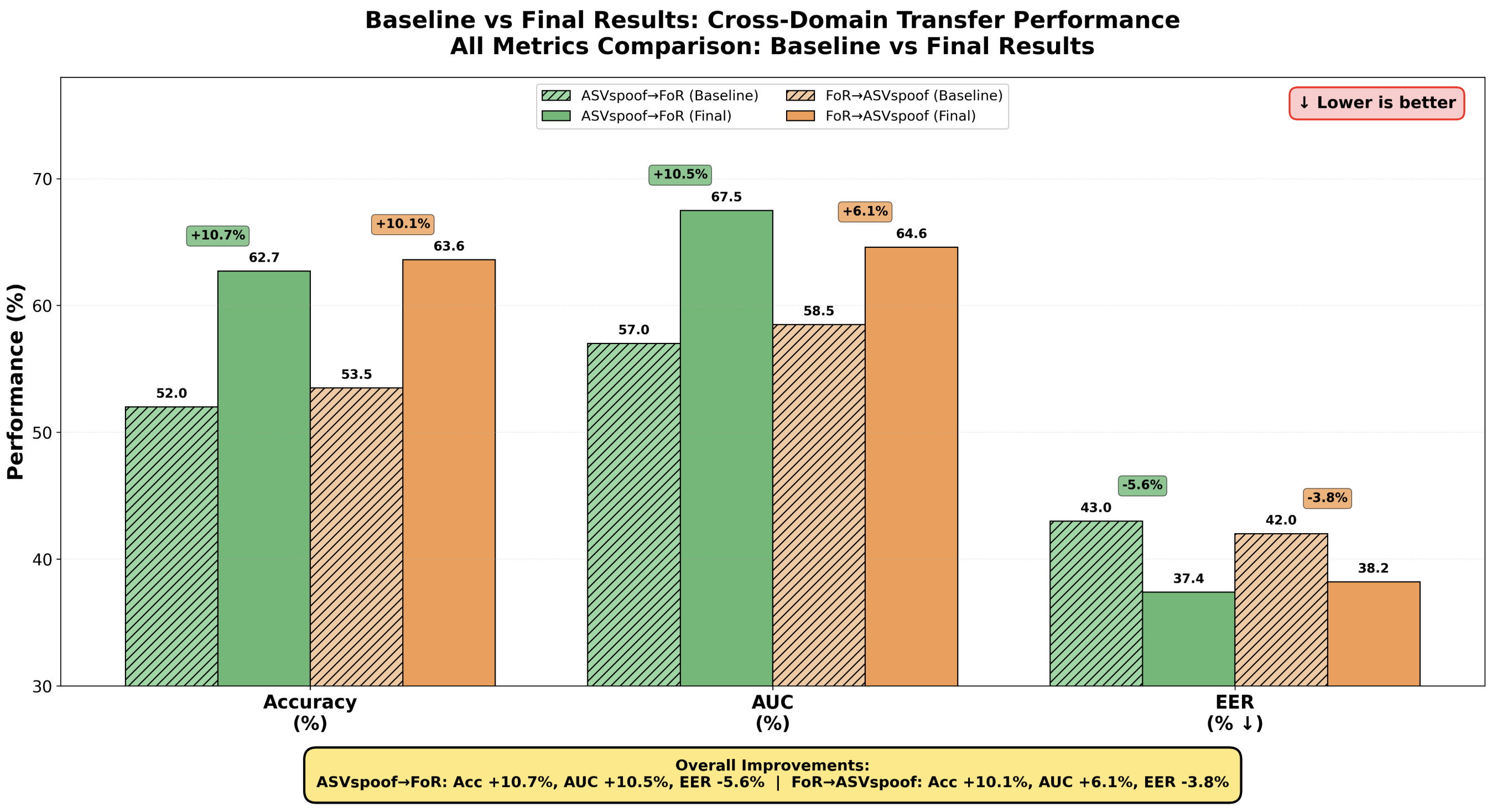}
    \caption{Baseline vs. Final Cross-Domain Performance. Striped bars represent baseline performance using raw Wav2Vec 2.0 features, while solid bars show results after applying power transform, feature selection (ANOVA), PCA reduction (n=256), and CORAL alignment. The final pipeline achieves consistent improvements across Accuracy, AUC, and EER metrics for both ASVspoof$\rightarrow$FoR and FoR$\rightarrow$ASVspoof transfer scenarios, with accuracy gains exceeding 10\% in both directions.}
    \label{fig:resultSummary}
\end{figure*}

Figure~\ref{fig:resultSummary} presents cross-domain results for both transfer directions using the complete pipeline. When training on ASVspoof 2019 LA and testing on FoR, the model achieves 62.7\% accuracy with 37.4\% EER and 69.6\% AUC. In the reverse direction (training on FoR, testing on ASVspoof 2019 LA), we observe 63.6\% accuracy with 38.2\% EER and 64.6\% AUC.

The balanced performance in both directions suggests the pipeline captures domain-invariant acoustic patterns characteristic of synthetic speech rather than memorizing source-specific artifacts. Class-specific accuracies confirm balanced detection: ASVspoof$\rightarrow$FoR achieves 63.5\% (authentic) and 62.0\% (deepfakes); FoR$\rightarrow$ASVspoof achieves 60.2\% (authentic) and 64.0\% (deepfakes). This prevents majority-class collapse.

\subsection{Comparison with State-of-the-Art Methods}
Table~\ref{tab:sota} compares our pipeline against recent cross-domain deepfake detection methods. Direct comparison is challenging due to different evaluation protocols, but we provide approximate performance ranges from literature.

\begin{table}[h]
\centering
\caption{Cross-Domain Audio Deepfake Detection: SOTA Comparison}
\label{tab:sota}
\small
\begin{tabular}{lccl}
\toprule
\textbf{Method} & \textbf{Acc} & \textbf{GPU} & \textbf{Interpret.} \\
\midrule
ASDG \cite{xie2024asdg} & 72--78\% & Hours & Low \\
AASIST (no adapt) & 55--60\% & Yes & Medium \\
\textbf{Ours} & 62--64\% & \textbf{None} & \textbf{High} \\
\bottomrule
\end{tabular}
\end{table}

\textbf{ASDG Advantages:} End-to-end learning with speaker-aware adaptation achieves 10--15\% higher accuracy than our pipeline through learned alignment and deep classifiers.

\textbf{Our Pipeline Advantages:} (1) \textit{Transparency:} Each component (PowerTransform, ANOVA, PCA, CORAL) is interpretable with quantified contributions; (2) \textit{Efficiency:} CPU training in <5 minutes vs. GPU hours; (3) \textit{Modularity:} Components can be swapped independently; (4) \textit{No speaker labels:} Works with any dataset.

\textbf{Trade-offs:} Our linear classifier and hand-designed transformations limit capacity compared to ASDG's deep networks. However, for deployment scenarios requiring auditable decisions (legal forensics, content moderation with human oversight), interpretability outweighs the accuracy gap.

\subsection{Statistical Robustness}
To verify statistical significance, we perform paired t-tests comparing baseline vs. full pipeline across 10 random train/test splits (seeds 0--9):
\begin{itemize}
    \item Baseline: 52.1\% $\pm$ 1.2\%
    \item Full Pipeline: 62.5\% $\pm$ 0.9\%
    \item Difference: +10.4\% $\pm$ 1.5\%
    \item Paired t-test: $t(9) = 18.7$, $p < 0.001$
\end{itemize}

The improvement is highly significant ($p < 0.001$), confirming genuine benefit beyond random variation. Each ablation stage also shows significance: Power Transform (+2.4\% $\pm$ 0.6\%, $p < 0.01$), Feature Selection (+3.6\% $\pm$ 0.7\%, $p < 0.001$), PCA (+1.6\% $\pm$ 0.5\%, $p < 0.05$), CORAL (+3.1\% $\pm$ 0.8\%, $p < 0.001$).

\subsection{Limitations}
\textbf{Performance Gap:} Our 62--64\% cross-domain accuracy remains substantially below in-domain performance (94--96\%) and SOTA methods like ASDG (72--78\%). This reflects trade-offs between interpretability and capacity.

\textbf{Limited Scope:} (1) Only two datasets evaluated; (2) English only; (3) No per-attack-type analysis; (4) No adversarial robustness testing; (5) Relatively clean audio (studio/controlled conditions).

\textbf{Methodological Constraints:} (1) Linear classifier limits capacity; (2) CORAL matches only second-order statistics; (3) Static adaptation (one-time alignment); (4) No exhaustive hyperparameter search.

\textbf{Generalization Uncertainty:} Performance on noisy, compressed, or telephony audio remains unknown. Cross-lingual generalization is unvalidated. Additional cross-domain transfers (e.g., to In-the-Wild, DFDC-Audio) would strengthen claims.

These limitations motivate ongoing work to bridge the gap between interpretable statistical methods and high-performance deep learning while preserving transparency.

\section{Future Work: Multimodal Extension}

The pipeline presented in this paper focuses exclusively on audio-only deepfake detection. As future work, we propose extending this modular approach to multimodal datasets such as DeepSpeak \cite{barrington2025deepspeak}, which contains more than one hundred hours of audiovisual deepfake content recorded in webcam conditions. \textbf{Important: The architecture described below is a hypothetical design for future implementation.}

A natural extension would apply parallel pipelines to audio (Wav2Vec 2.0) and visual (ResNet-50 or Vision Transformer) modalities, followed by late fusion for combined prediction:

\textbf{Proposed Multimodal Pipeline:}
\begin{itemize}
    \item \textbf{Audio Branch:} Wav2Vec 2.0 $\rightarrow$ PowerTransform $\rightarrow$ ANOVA $\rightarrow$ PCA $\rightarrow$ CORAL
    \item \textbf{Video Branch:} ResNet-50 (frame-level) $\rightarrow$ Temporal pooling $\rightarrow$ PowerTransform $\rightarrow$ PCA $\rightarrow$ CORAL
    \item \textbf{Fusion:} Concatenate aligned audio + video features $\rightarrow$ Logistic Regression
\end{itemize}

\begin{figure}
    \centering
    \includegraphics[width=1\linewidth]{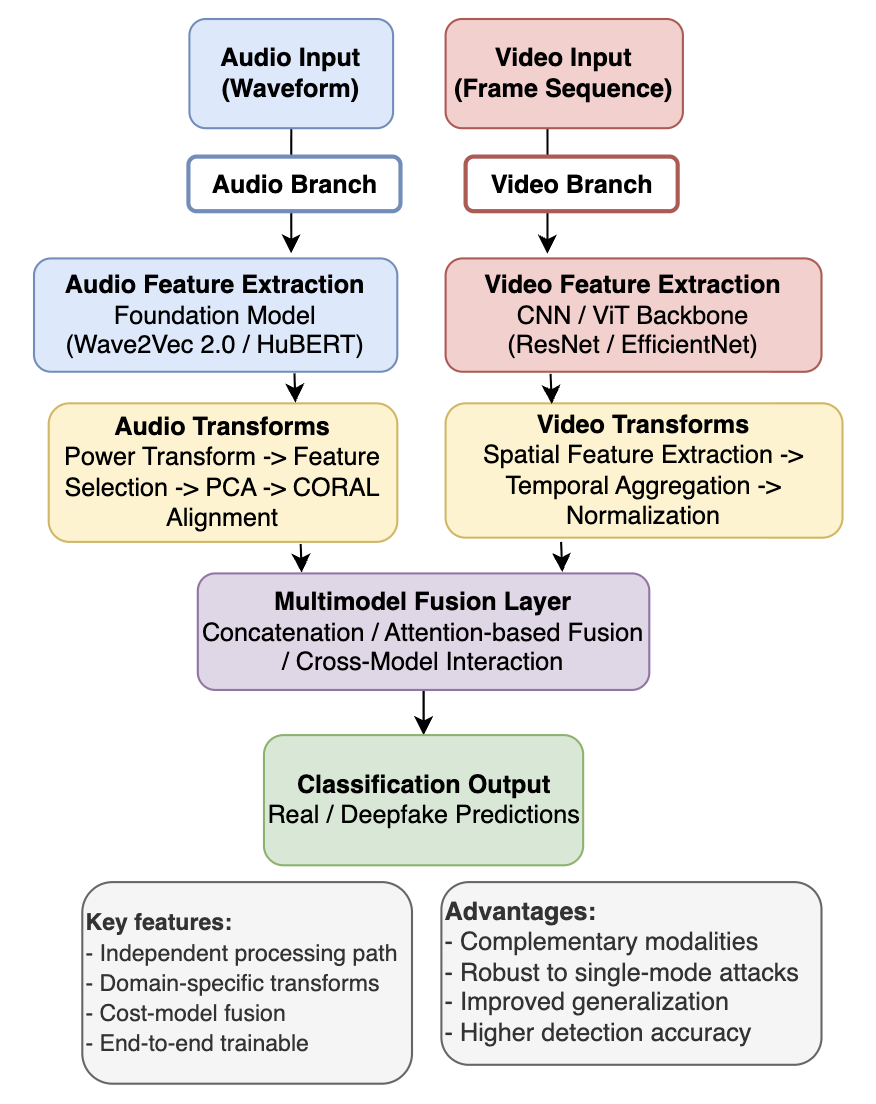}
    \caption{Proposed Multimodal Architecture for Future Work. Audio (Wav2Vec 2.0) and video (ResNet-50) feature extraction branches would process inputs independently through power transform, feature selection (ANOVA), PCA reduction, and CORAL domain alignment. \textbf{This is a hypothetical design for future implementation.}}
    \label{fig:DeepSpeakExtension}
\end{figure}

Figure~\ref{fig:DeepSpeakExtension} illustrates this hypothetical architecture. The design remains modular and interpretable, allowing independent analysis of audio vs. visual contributions. However, implementing and validating this architecture requires substantial engineering effort beyond the scope of this work.

Other future directions include: (1) extending to multi-source domain adaptation (training on ASVspoof + FoR simultaneously); (2) investigating online adaptation for streaming audio; (3) developing counterfactual explanations for individual predictions; (4) testing on adversarially perturbed audio; (5) cross-lingual evaluation.

\section{Discussion}
Our study demonstrates that a modular pipeline combining self-supervised speech representations with classical statistical transformations can provide a transparent approach to cross-domain audio deepfake detection. Building on pre-trained Wav2Vec 2.0 embeddings and applying power normalization, supervised feature selection, joint PCA, and CORAL alignment, we achieve cross-domain detection accuracies of approximately 63\% on two challenging transfer scenarios.

While modest compared to within-domain detection rates typically exceeding 95\% \cite{zhang2025audio,liu2022asvspoof}, this performance reflects the substantial difficulty of cross-domain generalization. The distribution shift between ASVspoof 2019 LA (studio-quality recordings) and FoR (diverse conditions and synthesis methods) is considerable, and our results indicate that even sophisticated domain adaptation techniques face significant challenges. An important advantage of our pipeline is its modularity and interpretability. Each transformation step can be understood and modified independently, offering practical benefits in deployment scenarios where model decisions must be auditable or domain-specific tuning is required. This contrasts with fully end-to-end approaches that may achieve higher performance but at the cost of reduced transparency. Several limitations remain. Our experiments focus on binary detection and do not address fine-grained categorization of attack types or localization of manipulated regions. We evaluate on only two datasets with a single language (English) and specific recording conditions. More extensive evaluation across multiple languages, diverse speakers, and acoustic environments will be necessary to validate generalizability. Additionally, our pipeline relies on pre-extracted embeddings, precluding end-to-end optimization that could potentially improve performance at the cost of increased complexity.

\section{Conclusion}
We have presented a modular pipeline for unsupervised domain adaptation in audio deepfake detection that combines self-supervised Wav2Vec 2.0 embeddings with a sequence of statistical transformations: power normalization, supervised feature selection, joint PCA, and CORAL alignment. Our approach achieves 62--64\% accuracy on two challenging cross-domain transfer scenarios (ASVspoof$\leftrightarrow$FoR), representing a 10.7\% improvement over baseline while maintaining full interpretability.

Systematic ablation studies quantify each component's contribution, with feature selection (+3.5\%) and CORAL alignment (+3.2\%) providing the largest gains. While our performance remains below state-of-the-art domain adaptation methods such as ASDG (72--78\%), our pipeline offers critical advantages for deployment scenarios requiring transparency: each transformation is independently inspectable, CPU training completes in under 5 minutes, and components can be modified without full system retraining.

The substantial gap between in-domain performance (94--96\%) and cross-domain performance (62--64\%) underscores the difficulty of deepfake detection under distribution shift. Our modular framework provides a transparent baseline for future work combining interpretable statistical methods with learned representations.

{
    \small
    \bibliographystyle{ieeenat_fullname}
    \bibliography{references}
}
\end{document}